\documentclass[12pt]{iopart}
\usepackage{palatino}
\usepackage{multirow}
\usepackage{graphicx}
\usepackage{appendix}
\usepackage{amssymb}
\usepackage{hyperref}
\usepackage{stmaryrd}
\usepackage{color}
\usepackage{url}
\usepackage[usenames,dvipsnames]{xcolor}
\usepackage{colortbl}

\begin{document}

\title[Environmental Influences on LIGO Detectors in S6]{Environmental Influences on the LIGO Gravitational Wave Detectors during the 6$^{th}$ Science Run}
\author{A Effler $^1$, R M S Schofield $^2$, V V Frolov $^3$, G Gonz\'alez$^1$, K Kawabe$^4$, J R Smith$^5$, J Birch$^3$, R McCarthy$^4$}
%Robert Schofield, Valery Frolov, Gabriela Gonzalez, Keita Kawabe, Joshua Smith, Jeremy Birch, Richard McCarthy}
\ead{aeffle2@lsu.edu, rmss@conch.uoregon.edu}
\address{$^1$Louisiana State University, Baton Rouge, LA 70803 USA,\\ $^2$University of Oregon, Eugene, OR 97403 USA,\\ $^3$LIGO Livingston Observatory, Livingston, LA 70754 USA, \\$^4$LIGO Hanford Observatory, Richland, WA 99354 USA, \\ $^5$California State University Fullerton, Fullerton, CA 92831 USA.}

\date{}
\maketitle
%\renewcommand{\abstract}{textbf{Introduction}}

%\tableofcontents

\begin{abstract}

We describe the influence of environmental noise on LIGO detectors in the sixth science run (S6), from July 2009 to October 2010. We show results from experimental investigations testing the coupling level and mechanisms for acoustic, electromagnetic/magnetic and seismic noise to the instruments. We argument the sensors' importance for vetoes of false positive detections, report estimates of the noise sources' contributions to the detector background,  and discuss the ways in which environmental coupling should be reduced in the LIGO upgrade, Advanced LIGO.
\end{abstract}

\section{Introduction}

\subsection{The Laser Interferometeric Gravitational-Wave Observatory}
The Laser Interferometric Gravitational-Wave Observatory (LIGO) is a network of ground based interferometric gravitational wave detectors that seeks to observe gravitational wave (GW) signals from astrophysical sources such as binary coalescence of neutron stars or black holes, supernova explosions, isolated spinning neutron stars and stochastic waves \cite{GenDet}. In 2005-2007 the detectors acquired data in the LIGO fifth science run (S5) with enough sensitivity to detect coalescence of binary neutron star systems at an average distance of 15 MPc \cite{dist}, an occurrence that is expected to happen only once every 50 years \cite{Rates}. The sixth science run (S6) started in July 2009 and ended in October 2010, with a small number of upgrades to test advanced LIGO technology, reaching a sensitivity of 20 Mpc. Two detectors were operated, one in Hanford, WA (LHO, the LIGO Hanford Observatory) and one in Livingston, LA (LLO, the LIGO Livingston Observatory). 
A typical GW strain sensitivity of LHO and LLO in S6 is shown in Fig. \ref{darms} \cite{s6sens}. In October 2010, the LIGO detectors involved in S6 were taken down to install Advanced LIGO detectors \cite{aLIGOgeneral}, a major upgrade to LIGO which is expected to detect GW signals from binary neutron star systems many times a year \cite{Rates}.

\subsection{The Detectors}
\label{introdet}
Each LIGO detector is a power-recycled Michelson interferometer with Fabry-Perot cavities in the 4km long arms \cite{GenDet}. With the exception of the laser source and some auxiliary beams for controlling the system, all hardware and laser beam paths are enclosed in ultra-high vacuum with an average pressure 3 10$^{-9}$ torr. The main fused silica optics which serve as the test masses for gravitational wave detection, are suspended by a single steel wire 0.36 mm in diameter attached to a suspension frame on a passive, multi-layered seismic isolation stack. The position of each optic is controlled with coil-magnet sensors/actuators.
%using five 5 mm by 2 mm magnets glued to each mirror (4 in a square formation on the back of the mirrors and one on the side). 

\begin{figure}[htp]
\centering
\includegraphics[scale=0.4]{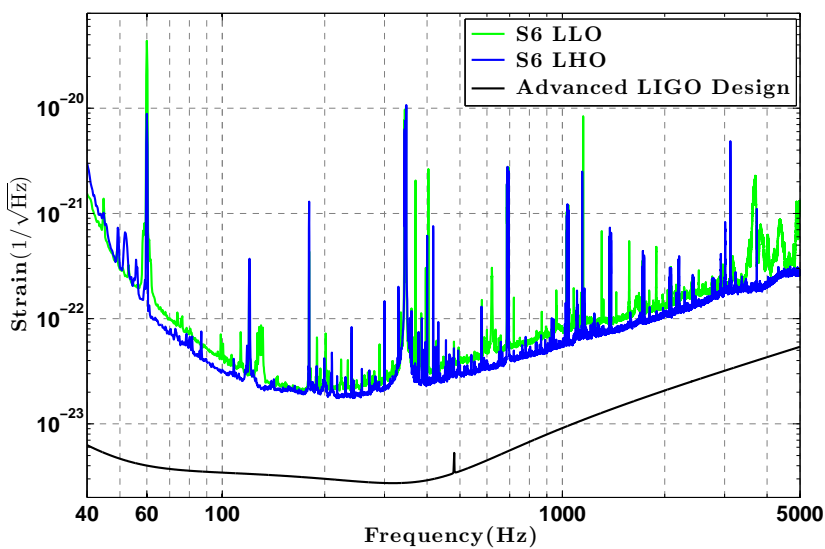}
\caption{\textit{Amplitude spectral density (ASD) of the noise limit to sensitivity for both detectors in S6 and design ASD for Advanced LIGO. The expected improvement in sensitivity for Advanced LIGO come from better optic suspensions and seismic isolation at low frequency, and from increased laser power and signal recycling (SR) methods at high frequency.}}
\label{darms}
\end{figure}

A GW will cause a signal on the detector manifested as a length difference between the two 4km arm cavities and measured as laser intensity fluctuations of the interference at the detector output, sometimes called the antisymmetric port \cite{SiggFreqResp}. This signal is digitized into a channel called DARM, short for "differential arm length". This channel can be calibrated \cite{S5Calibration} as dimensionless strain $\mathrm{h(t)=2*(L_x-L_y)/(L_x+L_y)}$, where h(t) is GW strain, $\mathrm{L_x}$ is the length of one arm and $\mathrm{L_y}$ is the length of the other arm. Alternatively, it can be expressed as displacement in meters: $\mathrm{\Delta L=L_x-L_y}$. The best sensitivity was achieved at 150 Hz is 3x10$^{-23}$1/$\mathrm{\sqrt{Hz}}$ x4km $\approx$ 10$^{-19}$m/$\mathrm{\sqrt{Hz}}$. A one-sided amplitude spectral density of this channel calibrated in unitless strain from each detector in S6 is shown in Fig.\ref{darms}.

Advanced LIGO will operate similarly, but with a significant number of upgrades. Also shown in Fig.\ref{darms} is the predicted design spectrum for Advanced LIGO at full power(at input $\sim$ 200W) with signal recycling. The estimated sensitivity to binary neutron star and black hole coalescence population is significantly higher for Advanced detectors due to the improved sensitivity below 100 Hz \cite{Rates}. Further improvements in sensitivity at higher frequencies due to lower thermal noise (due to the replacement of the optics) and lower shot noise (due to increased laser power) will give Advanced LIGO a thousand times the observable rate of initial LIGO.

\subsection{Environmental Effects}
The LIGO detectors have been carefully isolated from external non-astrophysical influences. Nevertheless, environmental disturbances can cause temporary or stationary contamination of the readout signal and reduce the sensitivity of the detector. 

For this reason, the LIGO instruments are equipped with dedicated sensors to detect environmental noise such as seismic, acoustic, magnetic or radio frequency electromagnetic disturbances. Typically these sensors are referenced to the DARM signal by creating or simulating environmental excitations large enough to be measured by both the environmental sensors and the DARM. The sensors provide critical information for the search of potential astrophysical events as well as for making the LIGO detectors less prone to environmental noise coupling.

Firstly, environmental sensors are used to validate astrophysical events by vetoing false positive GW signals \cite{S6DQ}. We must show that any candidate GW signal is extremely unlikely to have been produced by some environmental disturbance randomly coincident at both sites, or by a large scale effects affecting both detectors. During S6 (and before), for unmodeled burst searches,Êrandom coincident event triggers were frequent enough to be of concern.

Furthermore, we assess the level of the background environmental noise with respect to contaminating the DARM signal, typically in narrow frequency bands. Once the source and coupling mechanism is determined, we need to either remove the noise sources (e.g. shut down loud fans), attenuate the signal that propagates from the source to the coupling point on the detector (e.g. enclose the sensitive part of the detector in an isolation enclosure), or reduce the coupling at the detector (e.g. increase the size of the optics to avoid beam clipping). If we find a particularly high coupling location, we then add further monitors or relocate existing ones in order to ensure full validation of candidate astrophysical signals.

%The environmental monitors also help to improve the sensitivity of the interferometer when they show background environmental signals strong enough to dominate the noise in the GW channel over some narrow frequency band. In these cases we need to either remove the noise sources (e.g. shut down loud fans), attenuate the signal that propagates from the source to the coupling point on the detector (e.g. enclose the sensitive part of the detector in an isolation enclosure), or reduce the coupling at the detector (e.g. increase the size of the optics to avoid beam clipping). In addition, injections of environmental signals show where the sensitive parts of the interferometer are so that we can instrument it appropriately for validation of astrophysical events.
%
%We show that the LIGO environmental monitors are more sensitive to environmental signals than the gravitational wave channel. Our sensors are designed to detect any external event that could mimic a GW signal. To show this we create or simulate environmental excitations large enough to be measured by both the environmental sensors and the GW channel. Using these results we calculate the coupling of the ambient environmental noise and use the appropriate environmental sensors to diagnose the times at which the environmental effects may be too large in order to veto false positives \cite{S6DQ}.

In this paper we describe the main environmental influences, their effect on the detector and present a subset of studies representative of the efforts \cite{RobertPage} to reduce the environmental noise contribution to the detectors' background. In Section \ref{Env} we categorize the environmental influences by their coupling mechanisms into seismic, acoustic, audio frequency magnetic and radio frequency electromagnetic effects. For each class we discuss the possible sources, general mitigation approaches and the influence on the detector. In Section \ref{Meas} we show a set of studies performed in S6 characterizing each of these categories. In Section \ref{End} we discuss the results of the studies, how these methods will be used in the future and the implications for the more sensitive Advanced LIGO detectors.

\section {Characterizing the Detectors' Physical Environment} 
\label{Env}
The main categories of environmental influences on the detectors are seismic, acoustic and magnetic/electromagnetic field disturbances. We use accelerometers and seismometers to measure seismic motion, microphones to measure acoustic noise, magnetometers to measure audio frequency magnetic fields, radio receivers to track RF fields, temperature sensors to track temperature changes and voltage monitors to track the voltage of the electric power supplied to the site. Table 1 shows the details of some of the sensor types used in the LIGO Physical Environmental Monitoring (PEM) sub-system. 

\begin{figure}[htp]
\centering
\includegraphics[scale=0.7]{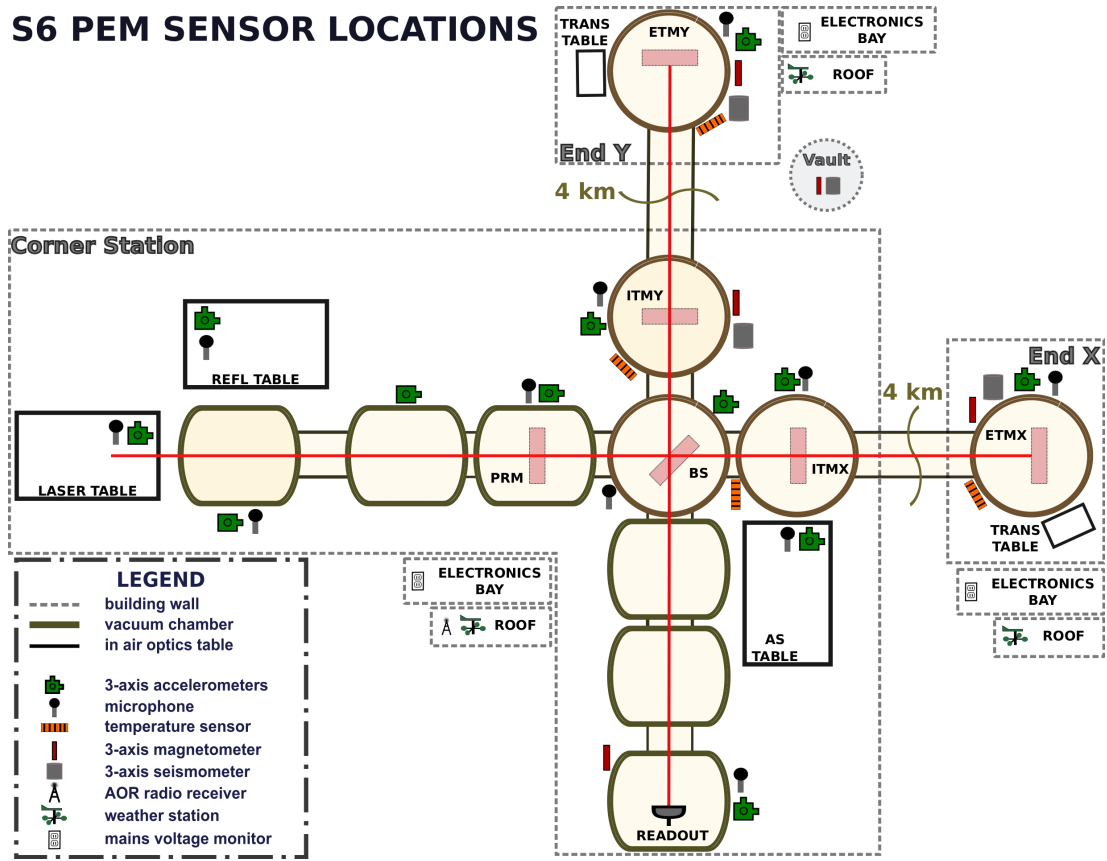}
\caption{\textit{The Physical Environmental Monitoring system layout at the LIGO Livingston detector during S6. The setup for LIGO Hanford was very similar. Shaded regions indicate the vacuum enclosure. Circles and rectangles indicate vacuum chambers where mirrors were suspended. Optical tables were surrounded by acoustic enclosures but were not in vacuum.}}
\label{llopem}
\end{figure}

Fig. \ref{llopem} shows the PEM sensor number and locations in S6 at LLO (with a similar setup at LHO). The reflection (REFL) table, the anti-symmetric (AS) table, the laser table and the transmission (TRANS) tables are not in vacuum and contain important feedback sensors for interferometer control. 

\begin{table}
\centering
\begin{tabular}{|c|c|c|c|c|} \hline
Type & Sensor & Operating Frequency & Sampling Frequency\\ \hline \hline
seismometer & Guralp\tiny{$^{\textregistered}$} & 0.1-20 Hz & 256 Hz\\ \hline
accelerometer & Wilcoxon{\tiny{$^{\textregistered}$}} 731-207 & 1-900 Hz & 2048 Hz\\ \hline
microphone &  Br\"{u}el\&Kjaer{\tiny{$^{\textregistered}$}} 4130 & 10-900 Hz  & 2048 Hz\\ \hline
magnetometer & Bartington{\tiny{$^{\textregistered}$}} 03CES100 & 0-900 Hz & 2048 Hz\\ \hline
radio station & AOR{\tiny{$^{\textregistered}$}} AR5000A & 24.5 MHz & 2048 Hz\\ \hline
\end{tabular}
\caption{Important PEM sensor types and the frequency ranges in which they are used. The frequency range is a combination of sensor calibration range from the manufacturer and the sampling rate at which they are recorded.}
\end{table}

LIGO is designed to not be dominated by environmental noise at frequencies higher than approximately 50 Hz, the frequency band of interest for GW in initial LIGO. Most control systems of the interferometer must operate at low frequencies and can become unstable due to large environmental disturbances. If these disturbances are too large, the interferometer will not be operational because optical cavities cannot be kept on resonance due to limited dynamic range (i.e. the detector is not "locked"). The interferometer is a gravitational wave detector only in its linear operating regime, i.e. when all optical cavities are stably locked near resonance for long periods of time \cite{GenDet}. 

There are several ways in which environmental noise can couple into the detector readout. The most significant ways are: changing the length of optical cavities, causing laser beam jitter, modulating the path length of scattered light which then recombines with the main laser beam, and introducing frequency noise.  However, environmental effects are often non-linear and cannot be removed offline in data processing. 

\subsection{Seismic Influences}

LIGO seismic isolation systems very efficiently reduce noise above 10 Hz, but amplify noise at the resonances of the mass-spring isolation stages. Feedback control systems keep the Fabry-Perot cavities locked on resonance, and low frequency seismic motion is the main contributor to the length and angular control signals. Moreover, large relative motion between mirrors in-vacuum suspended and out of vacuum sensing photodiodes can generate large control signals, which can cause upconversion \cite{RyanFF}.

The calibrated detector noise shows a steeply descending curve at low frequencies, called the "seismic wall", due to the residual seismic noise attenuation by suspensions and seismic isolation, following roughly a slope of f$^{-10}$ due to the five layers of isolation of the main optics \cite{Shyang}.

\begin{figure}[htp]
\centering
\includegraphics[scale=0.4]{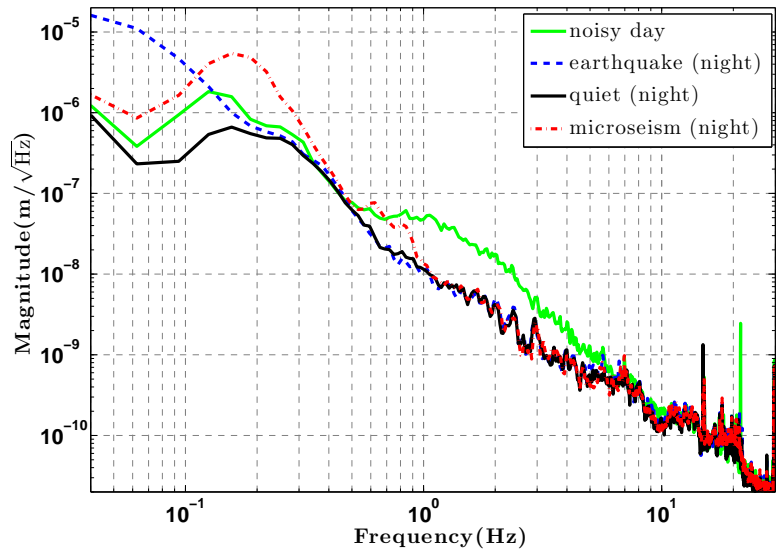}
\caption{\textit{The LIGO Livingston Observatory seismic background in different representative conditions, as seen by the horizontal axis of a seismometer located in the corner building. The "microseism" trace shows an instance where the oceanic microseism peak around 0.15 Hz was larger than average. The "noisy day" trace shows the effect of strong, nearby seismic disturbances due to human activity, in this particular case timber logging a few miles away from the detector. The "earthquake" trace shows an increased amplitude of seismic motion at very low frequency typical when large, far-away earthquakes occur. The "quiet" spectrum shows an instance of some of the quietest seismic environment we can expect.}}
\label{seismic}
\end{figure}

Transient sources of seismic motion include earthquakes, winds, ground and air traffic. The two detectors in Livingston, LA and Hanford, WA have different seismic backgrounds due to the very different geological structure of their locations \cite{Daw}.

Distant earthquakes produce ground motion with frequencies of 0.03 to 0.1 Hz, and even higher the closer the epicenter is to the detector. The interferometer seismic isolation is largely ineffective at these low frequencies, and many times the detector cannot remain locked. In Fig.\ref{seismic}  the "earthquake" curve shows the amplitude spectral density of a local seismometer signal during a 5.9 magnitude earthquake near Peru, with a peak around 0.05 Hz. Winds higher than 10-20 miles/hour will cause the buildings to sway enough to affect the detector output, even to the point where the detectors cannot remain locked. The effect of wind shows up mostly in the 0.5-15 Hz frequency range, but also as building tilt at lower frequencies.

%Vehicular traffic generally produces ground motion in the frequency band of 2-15 Hz. Because both sites are located away from cities but relatively close (2-4 miles) to large roads with truck traffic, there is a large difference between day and night operation. 

Vehicular traffic at highway speeds produce ground motion in the 2-15 Hz band (depending on axel spacing and speed) and distant human activities produce motion in the 1-3Hz range. Because the sources are anthropogenic, there is a large difference between day and night in these bands at both sites, despite their location away from cities. Other sources can show up in this frequency range as well, for example dam operations, forest logging or large scale construction.

Another source of seismic noise for the interferometers is storms in the oceans resulting in low frequency peaks generally highest at twice the wave propagation frequency, which called "microseism peaks" \cite{micro}. These are seen in the range of 0.07 to 0.7 Hz and couple directly to detector motion. All traces in Fig.\ref{seismic} show this peak around 0.15 Hz, but the one labelled "microseism" showcases a particularly high motion instance caused by storms in the Gulf of Mexico and detected in seismometers at LLO. This effect is always present, but may vary by up to two orders of magnitude on the time scale of a few days. Predictably, at LHO the Pacific storms have a larger influence. As with earthquakes, at these low frequencies there is little seismic attenuation, hence the motion couples directly to the detector's control systems. The seismometer signals were used to create a feedforward servo to reduce the coupling of seismic noise in this frequency band to the GW signal, with best results obtained mid-S6 when the coupling was reduced by a factor of 5 at the microseism with an overall RMS reduction factor of 2 on the DARM signal \cite{RyanFF}. 

Because the detectors have 4 km long arms and operate nearly continuously during science runs, earth tides caused by the gravitational pull of the Moon and Sun cause significant changes in the distance between the optics which would exceed the actuator range on the mirrors on a time scale of a few hours. To correct for this, the detectors have a tidal feedforward system which adjusts the position of the chambers or the laser frequency; it is based on tidal predictions calculated from the position of the Sun and Moon with respect to the detectors.

The previously described sources cannot be removed, so we reduce their coupling into the detector, e.g. by seismic feed-forward. A notable exception was the repaving of the main highway near the LHO site, which reduced truck traffic coupling into the detector by about a factor of two. Further, we characterize them well enough to be able to veto transients seen in the GW channel as non-astrophysical signals in our analyses \cite{Duncan}. 

There are also sources of vibration local to each observatory building such as motors, the air conditioning system, chillers and pumps, which cause seismic motion and disturb the detector's output. Many such sources have been localized and mitigated either by seismically isolating them on springs or changing their operation. 

\subsection{Acoustic Influences}

\begin{figure}[htp]
\centering
\includegraphics[scale=0.4]{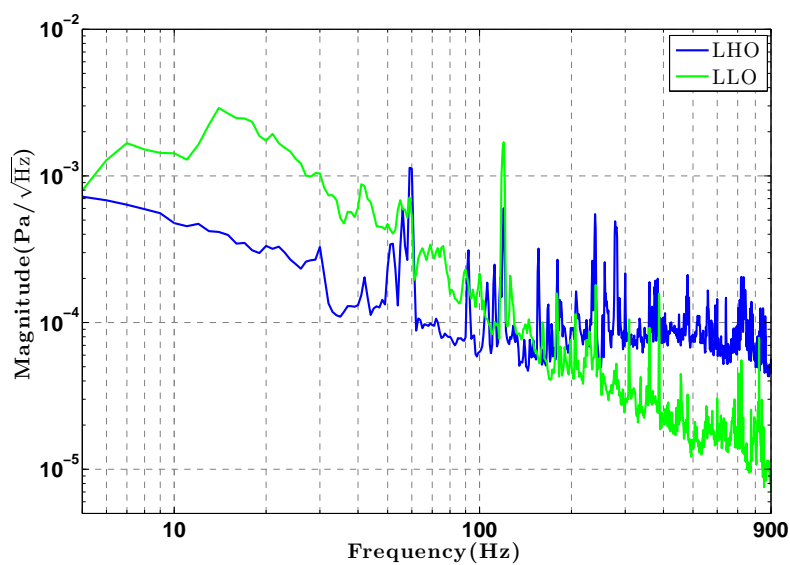}
\caption{\textit{Representative microphone spectra in the corner station at both sites, showing the acoustic background for the detector. At low frequency LHO has lower ambient acoustic noise due to extra insulation on the air conditioning system (to lower acoustic correlations between the previously present two colocated Hanford detectors). At high frequency LHO has a higher ambient acoustic noise due to the electronics racks which are in the same room as the detector, while at LLO the electronics racks are in a separate location. For Advanced LIGO the electronics will also be removed from the detector space at LHO.}}
\label{mics}
\end{figure}

Acoustic influences refer to sound waves propagating through air and vibrating components of the detector. Some known sources are electronics fans (above 50 Hz), chillers (below 60 Hz), building air control (below 100 Hz), thermally induced building creeks and thumps (broad band), nearby vehicles (50 - 150 Hz), and wind (broad band) \cite{RobertPage}. Propeller driven aircraft produce acoustic vibration in the range of 50-100 Hz if they fly close to the detector. A software monitor using data from microphones is used to veto such events seen in the GW channel \cite{PlaneMon}. Fig. \ref{mics} shows ambient normal spectra of the microphones in the corner station at each site. 

The in-vacuum systems of the interferometer including the photodiodes which read the GW signal are isolated to various degrees from direct acoustic wave propagation. However, several auxiliary systems are not in vacuum (see optical tables in Fig. \ref{llopem}) and have been found to be the major sources of coupling acoustic noise into the readout of the detector. For this reason, all out of vacuum optical tables have been acoustically insulated with enclosures in order to minimize the propagation of acoustic noise. 

Furthermore, acoustic noise can vibrate the outer suspension points of the in-vacuum system which then couples to high-frequency resonances of the seismic isolation. This effect has to be taken into account for auxiliary in-vacuum optics, especially those pertaining to the readout port. 

Acoustic noise has been found to couple primarily through beam jitter, beam clipping or scattering, all of which transform acoustically driven motion of the optical mounts into modulations on the primary and auxiliary photodiode signals. We have taken careful measures to reduce scatter, such as using beam dumps for stray beams, removing windows from all photodiodes, setting lenses at an angle and using damping material on optical mounts. 

\subsection{Magnetic Influences}

\begin{figure}[htp]
\centering
\includegraphics[scale=0.4]{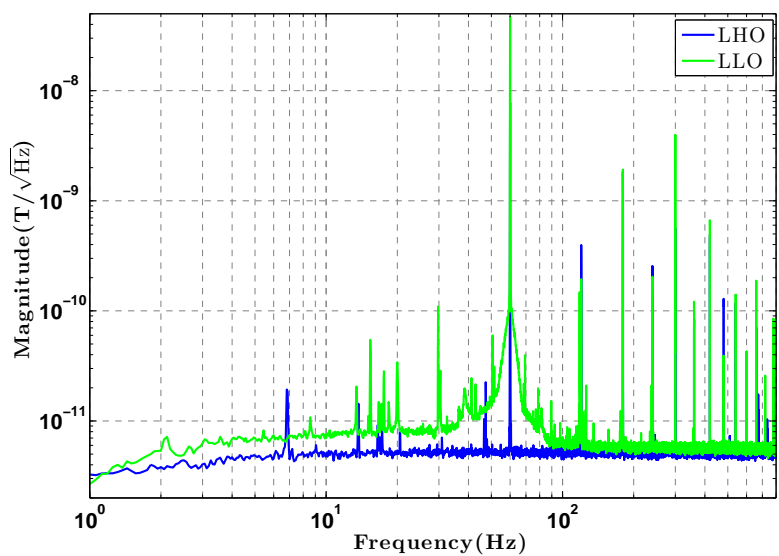}
\caption{\textit{Representative magnetometer spectra in the corner station at both sites, showing the magnetic background for the detectors. The Livingston detector has a higher magnetic background, likely due to power lines present much closer to the site than at the Hanford detector site. Except for the 60 Hz harmonics, both background levels are low enough that they are not a significant limit to LIGO sensitivity, as described in Section \ref{Meas}. The broadening of the 60 Hz peak in the LLO spectrum is due to glitches in the 60 Hz amplitude, whose source we have not yet located.}}
\label{mags}
\end{figure}

Magnetic noise sources relevant for LIGO are all of electric origin, such as building heaters, large motors, lights or relatively near-by high-voltage power lines up to 4km away from the site. These peaks are not stationary in frequency or amplitude, so they create noise in a wider frequency range than just a narrow peak at the respective frequency. Furthermore, since the 60 Hz harmonics from AC power lines are large in the GW signal as well as some of the control signals, other narrow-frequency peaks in the region of 0.1 to 10 Hz which are not in the LIGO frequency band of interest can modulate the GW signal producing sidebands around large peaks like the 60 Hz harmonics. Fig. \ref{mags} shows the ambient spectra of the magnetic background for the LIGO detectors in S6 in the corner station. 

The major coupling mechanisms for magnetic and electromagnetic noise involve electronic modules, cables and the magnets located on the interferometer optics (used as actuators for mirror position control). Each optic has four magnets glued on its back and one on the side surface, which are then actuated by coils mounted on the frame surrounding the optic. The magnets alternate in polarity so that uniform magnetic field gradients do not directly have a displacement effect on the optic (up to the level the four magnet strengths were matched). However, any magnetic field gradient comparable to the magnetic field produced by the actuation coil can introduce noise in the length measurement directly related to the detector output. 

The Crab pulsar \cite{Crab} would produce continuous GWs near 59.6 Hz (double its rotation frequency), which happens to be very close to the US electric power system frequency at 60 Hz. The possibility of detection or of a significant bound on the amplitude of its GW emission hinges on being able to take sensitive data for long periods of time, to resolve the coherent signal expected from a pulsar. Transients in current flow (such as motors turning on and off) reduce the sensitivity of the detector to the Crab pulsar by broadening the 60 Hz peak. For example, an unknown source of 60 Hz transients at LLO made the Crab pulsar search less sensitive than the LHO search by a significant factor of a few \cite{RobertPage}.

\subsection{Radio Frequency Influences}

LIGO uses a modulation-demodulation scheme, known as heterodyning, to generate error signals for controlling the length and angular degrees of freedom of the interferometer. In S5 it was found that radio frequency (RF) noise from the environment  could couple to the modulation frequencies in the interferometer and produce noise in the output signal in the frequency band of interest for GWs. A major change in the upgrade from the fifth science run to the sixth science run was changing the readout from heterodyne detection to a DC homodyne scheme \cite{aLIGOpath} \cite{DCreadout}. The DC readout scheme for the main differential arm length signal used in S6 should have reduced this coupling, but it is important to continue to monitor this coupling since RF modulation is still used in the auxiliary length and alignment control system.

\section{Injection Methods}
\label{Meas}
To quantify the effect that environmental influences have on LIGO sensitivity it is necessary to measure both the noise and its coupling to the detector output. Here we describe a method of environmental injections which quantifies the coupling by injecting an environmental signal with known amplitude and spectral content and measuring its effect on the detector. 

For seismic injections we use a weighted cart and shakers, for acoustic injections we use a large 500W speaker, for magnetic injections we use a 1m diameter, 100-turn copper coil and for RF injections we use an RF source far outside the buildings, set to 100 Hz near the main modulation frequency of our controls. We use a different power source than the one used for the detector electronics such that the current draws of our equipment do not couple through. For all measurements except the RF ones we place the noise source in the same room as the detector, trying to get large but equal distances both to the assumed coupling sites and the witness sensor. This presents some technical difficulty, since the coupling sites may not be known in advance. We perform injections in various locations with respect to the detector in order to locate the largest coupling point and understand the measurement errors. This is not always possible and comparisons of coupling factors calculated for different injection positions suggest that the error in coupling factors can be as high as a factor of two.

The coupling function is the ratio of the calibrated environmental signal amplitude to the resulting differential arm length displacement. We choose an amplitude large enough that an effect can be produced and measured in the detector output. Because the LIGO detectors are well isolated, the limit to injection amplitudes is most often set by saturation in the sensor readout rather than excessive excitation in the GW readout. The injected environmental signal is typically a harmonic comb produced by a ramped sawtooth waveform. At each frequency multiple we divide the signal amplitude in GW readout by the amplitude in the sensor signal, in calibrated units, giving us a coupling function. 

To estimate the level at which the ambient environmental noise couples into the detector, we multiply the measured coupling function by the normal ambient spectrum of the sensor. If this estimated background is an order of magnitude or more beneath the GW readout spectrum, we can say that for conditions close to those measured, this effect is not a significant or limiting noise source. We also track the coupling of these influences over long periods of time to look for variations and identify unwanted changes. 

A different method which has proven useful in the past is that of burst injections, which we briefly describe here. We apply a transient vibration to various locations around the detector, e.g. tapping an optical table, and look for large coupling sites. This method is hard to quantify due to variation in injection strength at the test location and closest relevant sensor, but in terms of relative effects on the GW signal we can narrow down and investigate suspected coupling sites.

Mitigation of either the source or the coupling is not always possible. Hence it is important to characterize and track the effect in question so as to introduce effective vetoes in the data for non-astrophysical events and quantitatively understand the limiting factors to the detector sensitivity \cite{S6DQ}.

\subsection{Seismic Studies}

Due to the strong attenuation provided by the LIGO seismic isolation, seismic noise limits LIGO's sensitivity at relatively low frequencies (under 50 Hz), below the band of best sensitivity around 100-200 Hz. However, nonlinear upconversion processes have the unwanted effect of converting low frequency seismic motion into noise in the frequency band of interest. Efforts to understand the mechanisms of upconversion have implicated Barkhausen noise in the magnets glued to the test masses themselves, or in the magnetic parts associated with the actuation on the mirrors \cite{S6DQ} \cite{weiss}  .

\begin{figure}[htp]
\centering
\includegraphics[scale=0.35]{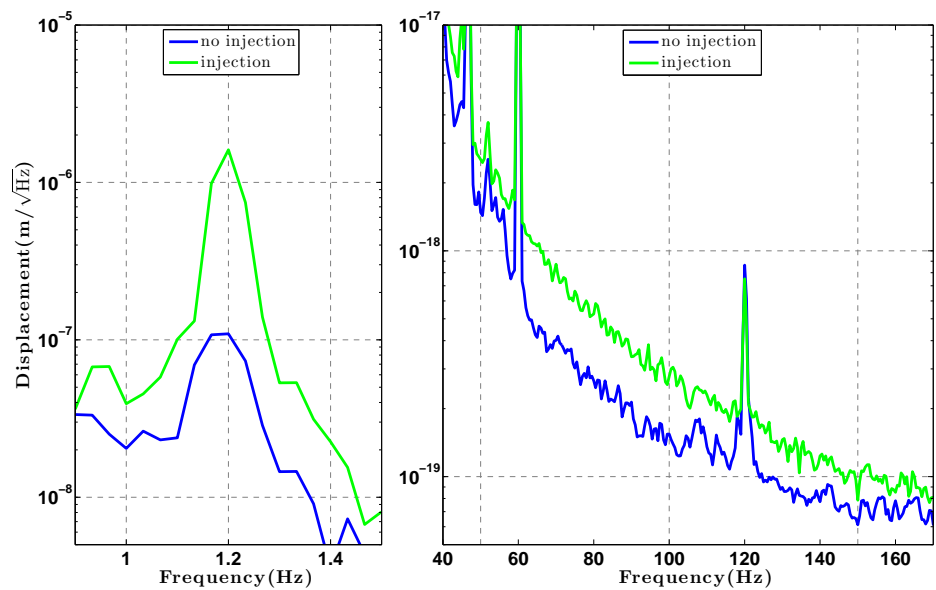}
\caption{\textit{LHO GW readout displacement showing upconversion of a 1.2 Hz seismic injection. Both panels show the same data of the displacement in the GW channel, but in different frequency bands. The left panel shows the linear effect of the injected signal in the detector output, while the right panel shows that this injection produces noise at higher frequencies (known as "upconversion").}}
\label{upconv}
\end{figure}

Seismic injection studies were performed differently at the two sites because the higher ambient ground motion at LLO required the implementation of a supplemental active seismic isolation system. At the Hanford site we use a weighted cart moved back and forth at a frequency of 1.2 Hz. We are able to see an effect in the detector output both at the injected frequency and at higher frequencies, demonstrating upconversion. 

The 1.2 Hz frequency was chosen because it is at a resonance of one of the passive seismic isolation stacks, and hence even with a relatively small injection we can excite enough motion in the interferometer to see upconversion. Fig. \ref{upconv} shows the result of a 1.2 Hz injection an order of magnitude larger in the detector output than the usual level at 1.2 Hz. At LLO a signal injected with the same amplitude cannot be seen due to the extra seismic attenuation. 

\subsection{Acoustic Studies}

To study the effect of acoustic noise coupling into the detector output we use a 500W speaker to produce an injection, and one or more microphones to measure the amplitude of the acoustic noise it produces. We ensure that the sensor is stationary and that the sound level at the studied coupling points and the sensors is about the same. In these studies, the approximate amplitude of the injection is 75mPa/$\mathrm{\sqrt{Hz}}$.

\begin{figure}[htp]
\centering
\includegraphics[scale=0.38]{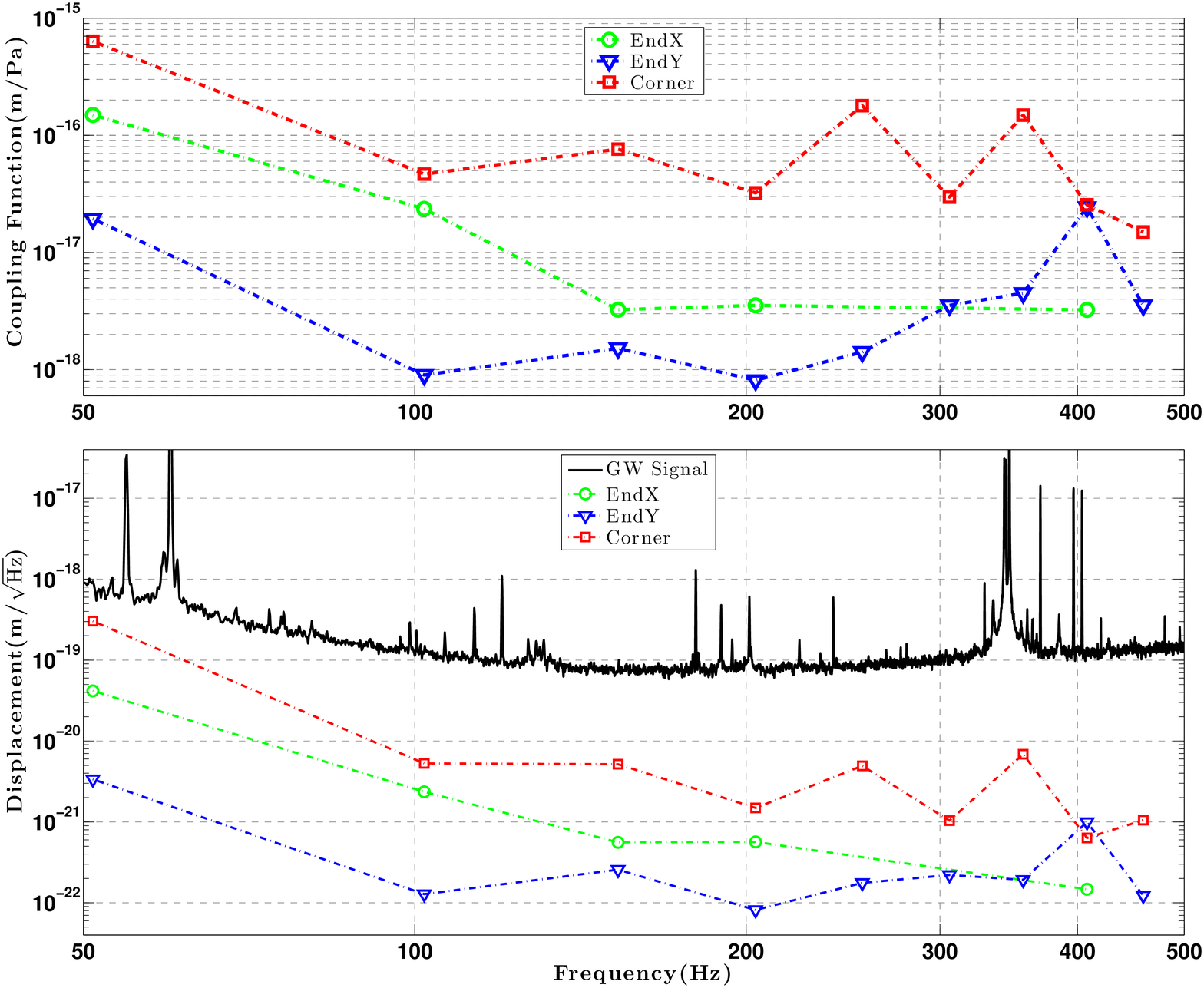}
\caption{\textit{Acoustic injection results at LLO for each detector building. The top panel shows the measured acoustic coupling function obtained from injecting acoustic noise in each builing. The calculated points then get multiplied by the ambient background level sensed by the microphones in each building to obtain the background estimate shown in the bottom panel.The coupling is only estimated at the injection frequencies marked in the plot.}}
\label{accf}
\end{figure}

\begin{figure}[htp]
\centering
\includegraphics[scale=0.38]{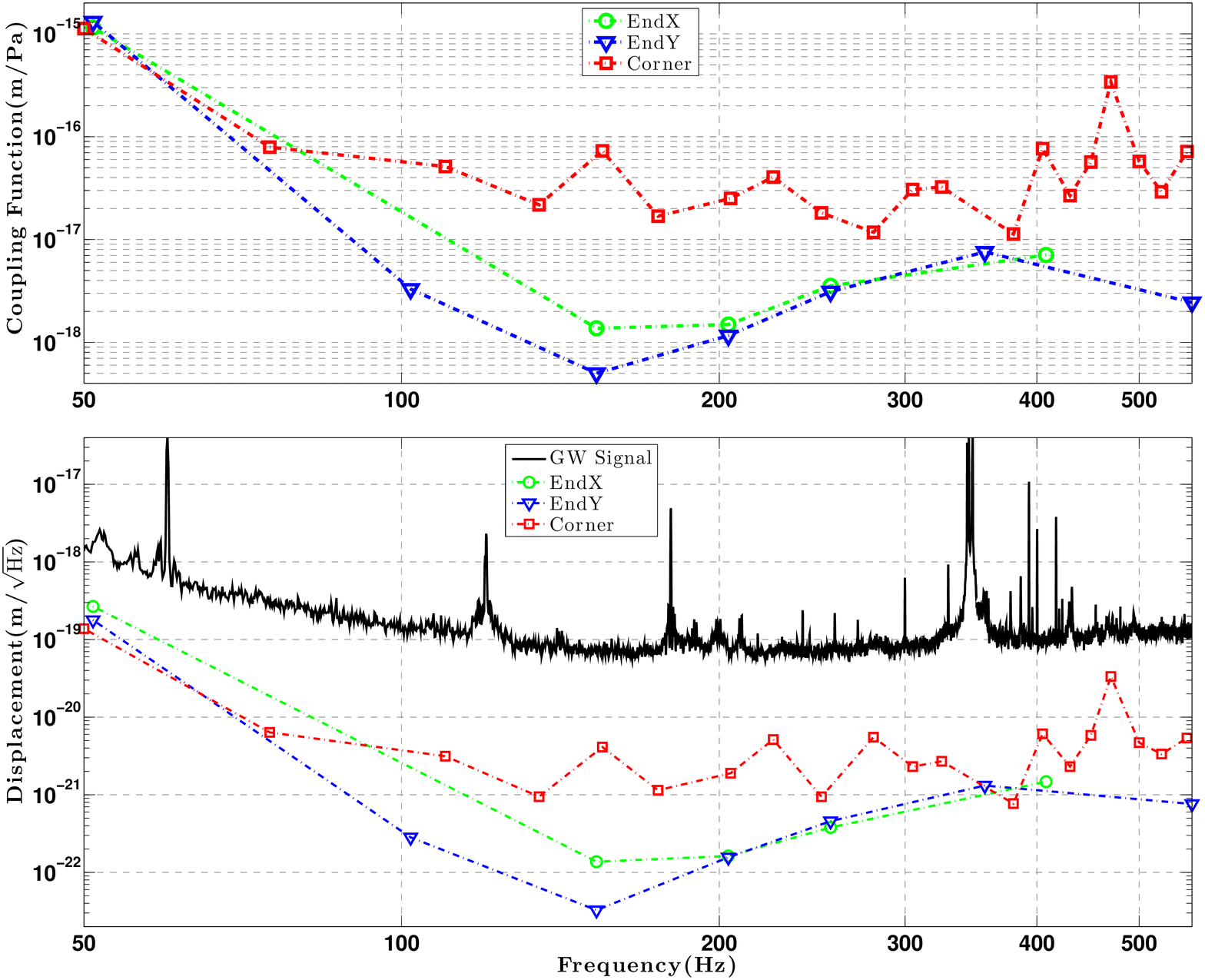}
\caption{\textit{Acoustic injection results at LHO for each detector building, equivalent to Fig. \ref{accf}.}}
\label{accf2}
\end{figure}

In the top panel of Fig.\ref{accf} and Fig.\ref{accf2} we show the measured coupling of acoustic noise to the detector output by taking the ratio of the signal seen in the detector output to the signal seen in the microphone. Then we multiply this coupling function by the normal ambient spectrum of the microphones to obtain a predicted contribution of ambient acoustic noise to the detector output. In the corner station the coupling is expected to be higher due to more of the detector subsystems and auxiliary control signals being present (see Fig. \ref{llopem}).

In the bottom panel of Fig.\ref{accf} we show the predicted ambient acoustic noise contribution at LLO and LHO respectively in S6 (measured in June 2010). At low frequencies the acoustic noise has a 20-30\% contribution to the noise in the detector output , while above 200 Hz it is more than 1 to 2 orders of magnitude lower than the strain noise and hence does not contribute significantly to the limit of the detector sensitivity. The amplitude variability is mostly due to the creation of nodes and antinodes of the sound waves in different locations around the detector room; we inject from different directions and placements to find the location from where the injection is the same in amplitude at the sensor and at the suspected coupling sites, but some mismatch remains. 

\subsection{Magnetic Studies}
To study magnetic noise coupling into the detector output we use a 1 meter diameter coil to create a magnetic field injection. We position the injection coil relatively far away (10-20 meters in the same room) such that the field produced would be the same at the studied coupling site (usually the magnet actuators on the optics) and the magnetometer. In these studies, the approximate injection amplitude used is 130nT/$\mathrm{\sqrt{Hz}}$.

\begin{figure}[htp]
\centering
\includegraphics[scale=0.38]{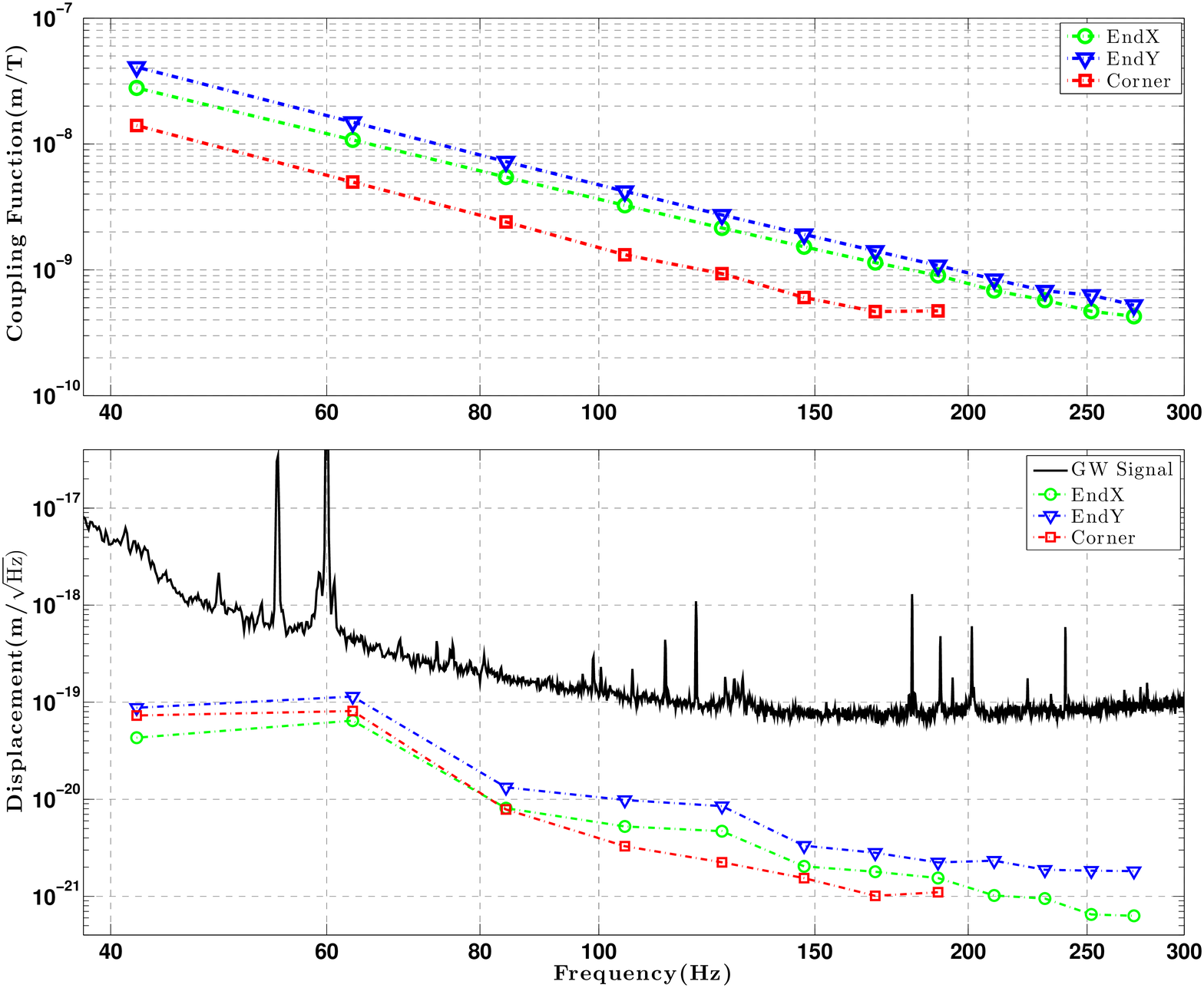}
\caption{\textit{Magnetic injection results at LLO for each detector building. The top panel shows the measured magnetic coupling function obtained from injecting magnetic noise in each building. The calculated points then get multiplied by the ambient background level sensed by the magnetometers in each building to obtain the background estimate shown in the bottom panel.The coupling is only estimated at the injection frequencies marked in the plot.}}
\label{magcf}
\end{figure}

\begin{figure}[htp]
\centering
\includegraphics[scale=0.38]{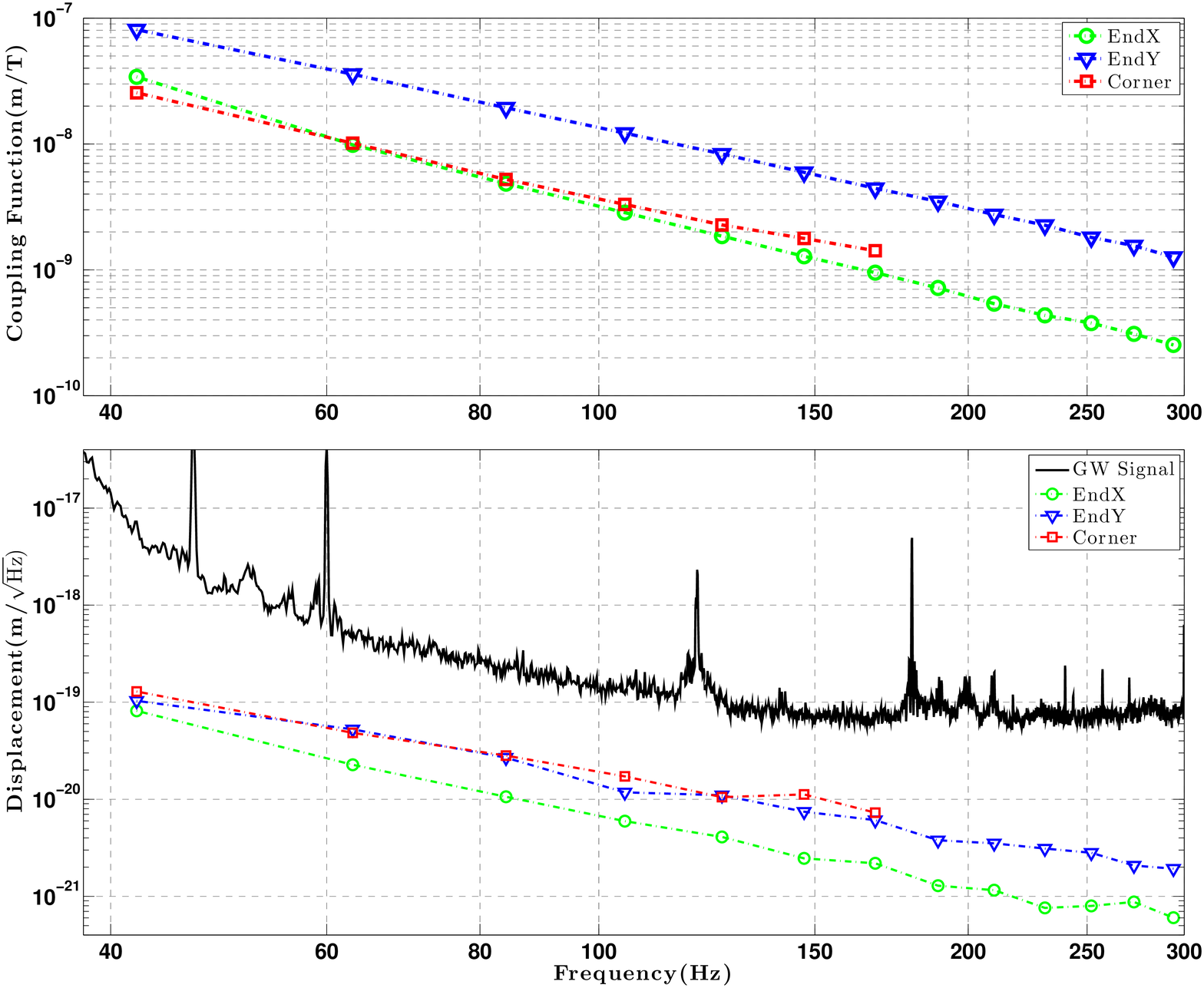}
\caption{\textit{Magnetic injection results at LHO for each detector building, equivalent to Fig. \ref{magcf}.}}
\label{magcf2}
\end{figure}

We use a calibrated magnetometer to read the size of the injections and compare to the signal amplitude in the detector output. This results in a magnetic coupling function, as the one shown for LLO in S6 in the top panel of Fig.\ref{magcf} and for LHO in Fig. \ref{magcf2} . We expect the coupling to depend on frequency as f$^{-3}$, with two factors of 1/$f^{2}$ from the pendulum response and one factor of 1/f from eddy current damping of the steel vacuum chamber.

The bottom panel of Fig.\ref{magcf} shows the predicted contribution of magnetic and electromagnetic ambient noise to the detector output. We conclude that, excluding the 60 Hz and its harmonics, ambient low-frequency electromagnetic noise did not significantly affect LIGO sensitivity in S6.

\subsection{Radio Frequency Studies}
\label{rfinj}

%\begin{figure}[htp]
%\centering
%\includegraphics[scale=0.4]{../Matlab/PaperFigs/Fig11.eps}
%\caption{\textit{LHO RF injections showing the change in RF coupling from the beginning of S6 to mid-S6, despite similar sensitivities at those two times. The 2009 trace shows some coupling into the detector, while the 2010 trace shows that with a larger amplitude than the previous measurement we were not able to produce a signal in the detector output. As can be seen in the bottom panel, we inject a signal several orders of magnitude larger than the background in both cases, and hence we can rule out this coupling as limiting the LIGO sensitivity.}}
%\label{RF}
%\end{figure}

To study radio frequency electromagnetic noise coupling into the detector output we place an amplified RF signal generator outside the corner building, far enough away (approximately 250m) to produce the same amplitude RF injection on the length scale of the detector corner components (where the RF control signals are read). We tune the injection close to 24.5 MHz, the main modulation frequency for the interferometer controls. We then measure the resulting injection strength with a radio antenna near the detector and compare to the signal we see in the detector readout.  

During the fifth science run, when RF readout was used, the RF ambient contribution was determined to be two orders of magnitude or more below the detector background. This was expected to become even lower with the use of DC readout, and was found to be 3 orders of magnitude below the detector output.

\section{Conclusions and Future Studies}
\label{End}

LIGO was designed to be isolated from environmental noise sources in its gravitational wave measurement band of 50 Hz to 7 kHz. Through injection studies we show that ambient stationary acoustic and magnetic sources do not contribute significantly to the noise limiting the S6 LIGO sensitivity. We have also shown the presence of seismic upconversion noise although the mechanism has not been fully understood. We prove that our PEM sensors are more sensitive to the environment than the detector output, making them essential in ruling out environmental causes for candidate GW signals. Furthermore, the PEM system has proven useful in investigating and eliminating undesired large couplings of acoustic, seismic and magnetic noise in various frequency bands.

The environment at the detector sites will remain the same for Advanced LIGO, but the vastly different design of Advanced LIGO will undoubtedly have very different coupling mechanisms and levels for the same environmental noise sources. The same level of environmental coupling as measured in S6 would however be a limiting noise source for Advanced LIGO and this has been carefully taken into account in the design of the seismic, suspension, optic actuation and other auxiliary systems. Advanced LIGO is not expected to be limited by environmental noise above 20 Hz.

With the experience gained from initial LIGO we will perform similar investigations for the Advanced LIGO detector, measuring the environmental coupling levels, searching for the causes of unwanted features or noise limits we discover in the new detector output, and mitigate any sources or mechanisms we find in the process. 

\ack
We thank the LIGO Laboratory and their staff without which this research would not be possible. We also thank Z Marka and N Christensen for useful comments. A Effler received support from the National Science Foundation through the grant 1205882.

%\bibliographystyle{unsrt}
%\bibliography{pemreferences.bbl}

\section*{References}
\begin{thebibliography}{10}

\bibitem{GenDet}
B.~Abbott et~al.
\newblock {LIGO: The Laser Interferometer Gravitational Wave Observatory}.
\newblock {\em Rep. Prog. Phys}, 72(7):076901, 2009.

\bibitem{dist}
J.~Abadie et~al.
\newblock {Search for gravitational waves from compact binary coalescence in
  LIGO and Virgo data from S5 and VSR1}.
\newblock {\em Phys. Rev. D}, 82(10):102001, 2010.

\bibitem{Rates}
J.~et~al Abadie.
\newblock {Predictions for the Rates of Compact Binary Coalescences Observable
  by Ground-based Gravitational-wave Detectors}.
\newblock {\em Class. Quant. Grav.}, 27(17):173001, 2010.

\bibitem{s6sens}
J.~Abadie et~al.
\newblock {Sensitivity Achieved by the LIGO and Virgo Gravitational Wave
  Detectors during LIGO's Sixth and Virgo's Second and Third Science Runs}.
\newblock non-journal companion to paper 4, LIGO and Virgo Scientific
  Collaboration, 2012.

\bibitem{aLIGOgeneral}
Gregory~M Harry and the LIGO Scientific~Collaboration.
\newblock {Advanced LIGO: The Next Generation of Gravitational Wave Detectors}.
\newblock {\em Class. Quant. Grav.}, 27(8):084006, 2010.

\bibitem{SiggFreqResp}
D.~Sigg et~al.
\newblock {Frequency response of the LIGO Interferometer}.
\newblock Internal document, t970084, LIGO, 1997.

\bibitem{S5Calibration}
J.~et~al Abadie.
\newblock {Calibration of the LIGO Gravitational Wave Detectors in the Fifth
  Science Run}.
\newblock {\em Nucl.~Instrum.~Meth.~A}, 624(1):223--240, 2010.

\bibitem{S6DQ}
LIGO~Scientific Collaboration.
\newblock Characterization of the ligo detectors during their sixth science
  run.
\newblock {\em in preparation, LIGO ref P1000142}, 2010.

\bibitem{RobertPage}
Robert Schofield.
\newblock {LIGO Environmental Influences\\
  \url{http://www.ligo-wa.caltech.edu/~robert.schofield/iLIGOenvironmentalInfl%
ueinces.htm}}.

\bibitem{RyanFF}
R.~DeRosa et~al.
\newblock {Global feed-forward vibration isolation in a km scale
  interferometer}.
\newblock {\em Class. Quant. Grav.}, 29(21):215008, 2012.

\bibitem{Shyang}
S.~Wen.
\newblock {\em {Improved Seismic Isolation for the Laser Interferometer
  Gravitational Wave Observatory with Hydraulic External Pre-Isolator System}}.
\newblock PhD thesis, Louisiana State University, 2008.

\bibitem{Daw}
E~J~Daw et~al.
\newblock {Long-term study of the seismic environment at LIGO}.
\newblock {\em Class. Quant. Grav.}, 21(9):2255, 2004.

\bibitem{micro}
R.~Cessaro.
\newblock {Sources of Primary and Secondary Microseisms}.
\newblock {\em B.S.S.A.}, 84(1):142--148, 1994.

\bibitem{Duncan}
D.~M.~Macleod et~al.
\newblock {Reducing the effect of seismic noise in LIGO searches by targeted
  veto generation}.
\newblock {\em Class. Quant. Grav.}, 29(5):055006, 2012.

\bibitem{PlaneMon}
E.~Goetz et~al.
\newblock {PlaneMon: Airplane Detection Monitor}.
\newblock Internal document, t050174, LIGO, 2005.

\bibitem{Crab}
B.~Abbott et~al.
\newblock {Beating the spin-down limit on gravitational wave emission from the
  Crab pulsar}.
\newblock {\em Astrophys J. Lett}, 683(1):L45, 2008.

\bibitem{aLIGOpath}
J.~Smith.
\newblock {The path to the enhanced and advanced LIGO gravitational-wave
  detectors}.
\newblock {\em Class. Quant. Grav.}, 26(11):114013, 2008.

\bibitem{DCreadout}
T.~T. {Fricke}, N.~D. {Smith-Lefebvre}, R.~{Abbott}, R.~{Adhikari}, K.~L.
  {Dooley}, M.~{Evans}, P.~{Fritschel}, V.~V. {Frolov}, K.~{Kawabe}, J.~S.
  {Kissel}, B.~J.~J. {Slagmolen}, and S.~J. {Waldman}.
\newblock {DC readout experiment in Enhanced LIGO}.
\newblock {\em Classical and Quantum Gravity}, 29(6):065005, March 2012.

\bibitem{weiss}
R.~Weiss.
\newblock {Various Reports of Experiments Conducted on the Barkhausen Noise
  Various Reports of experiments conducted on the barkhausen noise research}.
\newblock internal document, t080355, MIT, 2008.

\end{thebibliography}
\end{document}